\documentclass[pra,prl,twocolumn,showpacs,floatfix,10pt]{revtex4-1}
\usepackage{graphicx}
\usepackage{dcolumn}
\usepackage{bm}

\usepackage{amsmath}
\usepackage{latexsym,amssymb,graphicx}
\usepackage[mathscr]{eucal}
\usepackage[bookmarks=true,
   colorlinks=true,
   linkcolor=blue,
   urlcolor=blue,
   citecolor=blue,
   bookmarks=true,
   hyperindex=true
]{hyperref}

\begin{document}


\title{Constraining the generalized uncertainty principle with cold atoms}

\author{Dongfeng Gao\textsuperscript{1,2,}}
\altaffiliation{Email: dfgao@wipm.ac.cn}
\author{Mingsheng Zhan\textsuperscript{1,2,} }
\altaffiliation{Email: mszhan@wipm.ac.cn}
\vskip 0.5cm
\affiliation{1 State Key Laboratory of Magnetic Resonance and Atomic and Molecular Physics, Wuhan Institute of Physics and Mathematics, Chinese Academy of Sciences - Wuhan National Laboratory for Optoelectronics, Wuhan 430071, China\\
2 Center for Cold Atom Physics, Chinese Academy of Sciences, Wuhan 430071, China }

\date{\today}

\begin{abstract}
Various theories of quantum gravity predict the existence of a minimum length scale, which implies the Planck-scale modifications of the Heisenberg uncertainty principle to a so-called generalized uncertainty principle (GUP). Previous studies of the GUP focused on its implications for high-energy physics, cosmology, and astrophysics. Here, the application of the GUP to low-energy quantum systems, and particularly cold atoms, is studied. Results from the $^{87}$Rb atom recoil experiment are used to set upper bounds on parameters in three different GUP proposals. A $10^{14}$-level bound on the Ali-Das-Vagenas proposal is found, which is the second best bound so far. A $10^{26}$-level bound on Maggiore's proposal is obtained, which turns out to be the best available bound on it.

\end{abstract}

\maketitle

\section{Introduction}

Establishing a quantum theory of gravity (also called quantum gravity) is one of the main challenges in modern physics. To gain insights into the development of such quantum theories, it is useful to investigate experimentally accessible quantum gravity effects. In general, such effects are very small, expected to be inversely proportional to the Planck energy scale $E_P= \sqrt{\hbar c^5 /G}=1.2 \times 10^{19}$ GeV. Even so, current experimental results, mainly from high-energy physics, astrophysics, and cosmology, can set meaningful bounds on parameters relevant to models on quantum gravity. A lot of work has been done, and references can be found in a recent review paper \cite{camelia2013}.

Here, we are interested in one particular quantum gravity effect, the so-called generalized uncertainty principle (GUP). According to the Heisenberg uncertainty principle (HUP), the uncertainties in the measurement of the length and the momentum satisfy the relation $\Delta x \Delta p \geq |\langle[x,p]\rangle|=\hbar/2$. In other words, the uncertainty $\Delta x$ is bounded by $\Delta x \geq \hbar /(2\Delta p)$. Therefore, on the cost of the information of the momentum $p$, the length can be arbitrarily precisely measured. On the other hand, it has long been known that various models on quantum gravity predict the existence of a minimum measurable length \cite{maggiore1993,gross1987,amati1989,konishi1990}. Thus, this forces people to modify the HUP to the GUP.

The GUP has various implications on a wide range of physical systems, especially in the quantum regime. For example, effects of the GUP on the thermodynamics of the quark-gluon plasma were studied in Ref. \cite{elmashad2014}. The impact of the GUP on thermodynamical parameters and the stability of the Schwarzschild black hole was investigated in Ref. \cite{sabri2012}. The GUP approach was also used to calculate quantities of the inflationary dynamics and the thermodynamics of the early universe \cite{tawfik2013}. The violation of the equivalence principle due to the GUP effects was discussed in Refs. \cite{tkachuk2013,ghosh2013}. Furthermore, it was suggested that experimental data of the Lamb shift, the Landau levels, and the tunneling current in a scanning tunneling microscope could be used to constrain the GUP corrections \cite{das2008, das2011}. A direct measurement of the GUP effects in quantum optics laboratory was proposed in Ref. \cite{pikovski2012}. More references can be found in Ref. \cite{tawfik2014}.

In recent years, rapid technological advances in atom interferometry have been achieved. Due to their high sensitivity, atom interferometers have already been used
in various precision measurements, and many important experimental results came out. For example, the value of the Newtonian gravitational constant was measured to be $G=6.67191(99)\times 10^{-11} {\rm m}^3 \, {\rm kg}^{-1}{\rm s}^{-2}$ \cite{tino2014}, which just differs by 1.5 combined standard deviations from the CODATA recommended value \cite{codata2010}. A $10^{-8}$-level test of the weak equivalence principle (WEP) was reported in a recent work \cite{zhan2015}. Atom interferometers can also be used to search deviations from Newton's $1/r^2$ law at the micrometer scale \cite{tino2006}. Another important application of atom interferometers in precision measurements is the determination of the fine structure constant, $\alpha$. A value of $\alpha^{-1}$=137.035999037(91) was obtained \cite{clade2011}, which is the second best value compared to the one deduced from the electron anomaly measurement \cite{hanneke2008}. More details on atom interferometers can be found in many review papers, such as Ref. \cite{cronin2009}. Inspired by these achievements with atom interferometry, people began to think about the possibility of studying quantum gravity effects on atomic physics. One such study was given in Ref. \cite{tino2009}, where the authors used the atom recoil measurement \cite{wicht2002} to constrain parameters in the energy-momentum dispersion relation modified by quantum gravity.

In this paper, we investigate the possibility of using the cold atom recoil measurement to constrain parameters in three popular GUP proposals. Improved bounds on these parameters are found. The paper is organized as follows. In Sect. II, three popular GUP proposals are introduced. Then a brief description of the cold atom recoil experiment is given in Sect. III. Through a detailed calculation of the GUP effects on the cold atom recoil measurement, bounds on the GUP parameters are obtained in Sect. IV. Finally, we conclude the paper in Sect. V.

\section{Three proposals on the GUP}

As discussed above, various models on quantum gravity such as string theory and loop quantum gravity suggest the existence of a minimum measurable length. Consequently, this indicates that the HUP should be modified to the GUP at energies close to the Planck energy scale $E_P$. But at the moment no model has the power of making robust predictions on what the GUP should be. The alternative way is making various proposals on the GUP. Depending on what indications from models on quantum gravity should be incorporated, many different GUP proposals have been made \cite{tawfik2014}. Here, we discuss three popular ones.

First, let us consider the so-called Kempf-Mangano-Mann (KMM) proposal, which was first discussed in Ref. \cite{kmm1995}. The authors were motivated by the observation that a variety of models of quantum gravity predicted a leading quadratic-in-the-momenta-type correction to the HUP. To incorporate this observation, the following form is proposed:
\begin{equation}
[x_i, p_j]=i \hbar \left(\delta_{ij} + \frac{\beta_0}{(M_P c)^2} \delta_{ij} p^2 + \frac{2\beta_0}{(M_P c)^2} p_i p_j\right),
\label{kmm}
\end{equation}
where $\beta_0$ is a dimensionless parameter, and $M_P = \sqrt{\hbar c /G}$ is the Planck mass. All other commutation relations vanish. It is easy to find the following uncertainty relation
\begin{equation}
\Delta x_i \Delta p_i \geq \frac{\hbar}{2} \left(1 + \frac{\beta_0}{(M_P c)^2}\left((\Delta p)^2+\langle p \rangle^2 + 2 \Delta p_i^2 +2\langle p_i \rangle^2 \right) \right),
\end{equation}
where $p^2=\sum_{j=1}^3 p^j p_j$. This inequality relation implies a minimum measurable length $\Delta x_{min}=\sqrt{3\beta_0}L_P$, where $L_P=\sqrt{\hbar G /c^3}$ is the Planck length. Theoretically, the Planck length is believed to be the minimal measurable length. Thus, $\beta_0$ is normally assumed to be of the order of unity. However, if one does not take the above assumption \textit{a priori}, current experiments can set upper bounds on it. For example, the standard model of high-energy physics is well tested at energy scale 100 GeV, which implies that $\beta_0 \leq 10^{34}$. Better bounds are obtained in Ref. \cite{das2008}, where the best one $\beta_0 \leq 10^{21}$ is set by the tunneling current measurement.

Next, we discuss the Ali-Das-Vagenas (ADV) proposal, which was first put forward in Ref. \cite{das2009}. The authors observed that doubly special relativity theories \cite{camelia2002} suggested a leading linear-in-the-momenta type correction to the HUP. To incorporate both the linear and the quadratic in-the-momenta corrections, they proposed the following form:
\begin{eqnarray}
\nonumber [x_i, p_j]&=& i \hbar \left(\delta_{ij} - \frac{\eta_0}{M_P c} (\delta_{ij} p+ p_i p_j/p)\right.\\
 & & \left.  + \frac{\eta_0^2}{(M_P c)^2}(\delta_{ij} p^2+ 3 p_i p_j)\right),
\end{eqnarray}
where $\eta_0$ is a dimensionless parameter, and is normally assumed to be of the order of unity. All other commutators vanish. Again, if the assumption on $\eta_0$ is not taken \textit{a priori}, one can use current experiments to set upper bounds on it. The minimum measurable length of this proposal is easy to find to be $\Delta x_{min}=2\eta_0 L_P$. Then, an immediate bound from high-energy physics is $\eta_0 \leq 10^{17}$. The best bound is set by the Lamb shift measurement \cite{das2011}, which is $\eta_0 \leq 10^{10}$.

The last proposal to be discussed is Maggiore's proposal \cite{maggiore1993,maggiore1995}, which is motivated by the relationship between the GUP and the quantum deformation of the Poincar${\rm\acute{e}}$ algebra. The author gave the proposal
\begin{equation}
[x_i, p_j]=i \hbar \, \delta_{ij}\sqrt{1+\frac{\gamma_0}{(M_P c)^2}(p^2+m^2 c^2)},
\label{maggiore1}
\end{equation}
where $\gamma_0$ is a dimensionless parameter, and is normally assumed to be of the order of unity. If this assumption is not taken \textit{a priori}, current experiments can be used to set upper bounds on it. The minimum measurable length of this proposal is found to be $\Delta x_{min}\simeq\sqrt{\gamma_0/2} L_P$. Again, a direct bound from high-energy physics is $\gamma_0 \leq 10^{34}$. Currently, no other bounds are available.
Since $\gamma_0/(M_P c)^2$ is very small, one can make Taylor series expansion on Eq. (\ref{maggiore1})
\begin{equation}
[x_i, p_j]=i \hbar \, \delta_{ij}\left(1+\frac{\gamma_0}{2(M_P c)^2}(p^2+m^2 c^2)\right).
\end{equation}
Compared to the KMM proposal, the above formula is very similar to it, and can be regarded as a generalization of it. But the difference will be important later.

\section{The cold atom recoil experiment}

\begin{figure}[htbp]
\centering
\includegraphics [width=10.6cm]{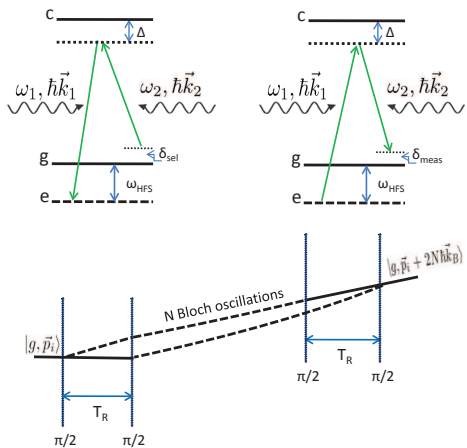}
\caption{Schematic diagram of Ramsey-Bord${\rm \acute{e}}$ type atomic interferometer combined with $N$ Bloch oscillations.}
\label{fig1}
\end{figure}

A typical cold atom recoil experiments can be described with the schematic diagram shown in Fig. 1, where two hyperfine ground states of atoms are labeled by $|g\rangle$ and $|e\rangle$, respectively. It combines a Ramsey-Bord${\rm \acute{e}}$-type atomic interferometer with Bloch oscillations so that as many as possible recoils can be coherently transferred to atoms and atomic velocities can be accurately measured with the Raman transition between two hyperfine ground states. Details of the experimental setup can be found in Ref. \cite{clade2006}. The experimental procedure can be described as follows.

A slow atomic beam, prepared in the $|g\rangle$ state by a two-dimensional magneto-optical trap (2D-MOT), is loaded into the interferometer through a standard three-dimensional magneto-optical trap (3D-MOT). After a few seconds of cooling, a first pair of Raman $\pi$/2 pulses, realized by using two counterpropagating laser beams with frequencies $\omega_1$ and $\omega_2$, and wave vectors $\vec{k}_1$ and $\vec{k}_2$, is applied to create two coherent atomic beams in the $|e\rangle$ state, and define an initial velocity distribution centered on $\vec{v}_i$.  Then, atoms remaining in the $|g\rangle$ state are blown away by a resonant laser beam tuned to the single photon transition. This process is called velocity selection of atoms. At resonance, according to the energy conservation law, we have
\begin{equation}
\hbar (\omega_1 - \omega_2 -\omega_{HFS})= \hbar\Delta + \frac{(\hbar (\vec{k}_1-\vec{k}_2)+m \vec{v}_i)^2-m^2 \vec{v}_i^2}{2 m} ,
\end{equation}
where $m$ is the mass of atoms, $\Delta$ is the single photon detuning of the atomic levels, and $\hbar \omega_{HFS}=E_g-E_e$ is the energy difference between the $|g\rangle$ state and the $|e \rangle$ state. This equation can be written as
\begin{equation}
 \delta_{sel}=\Delta + \frac{\hbar}{2 m}(\vec{k}_1-\vec{k}_2)^2+(\vec{k}_1-\vec{k}_2)\cdot \vec{v}_i,
\label{sel}
\end{equation}
where $\delta_{sel}\equiv \omega_1 - \omega_2 -\omega_{HFS}$ is defined to be the Raman detuning from the atomic resonance.

After the velocity selection, a number of $N$ Bloch oscillations are applied to both atomic beams. Bloch oscillations are an efficient way of transferring very high recoil velocities to atoms in a short time, while leaving the internal state of atoms unchanged. By using two counterpropagating laser beams, atoms are coherently accelerated through a succession of two-photon Raman transitions. In each Bloch oscillation, the atomic velocity increases by 2$\vec{v}_r$, where $\vec{v}_r=\hbar \vec{k}_B/m$ is the recoil velocity of the atom when absorbing a photon of momentum $\hbar \vec{k}_B$. Thus, a final velocity, $\vec{v}_f=\vec{v}_i + 2 N \vec{v}_r$, is achieved for atoms.

Finally, a second pair of Raman $\pi$/2 pulses is applied to read out the final velocity distribution centered on $\vec{v}_f$, and transfer the atoms back to the $|g\rangle$ state. Again, according to the energy conservation law, the Raman detuning for the velocity measurement is found to be
\begin{equation}
\delta_{meas}=\Delta + \frac{\hbar}{2 m}(\vec{k}_1-\vec{k}_2)^2+(\vec{k}_1-\vec{k}_2)\cdot \vec{v}_f .
\label{meas}
\end{equation}

Subtracting Eq. (\ref{sel}) from Eq. (\ref{meas}), one can easily find that
\begin{equation}
|\delta_{sel}-\delta_{meas}|= (k_1+k_2)|v_f-v_i|.
\end{equation}
Then, the ratio $h/m$ is determined by
\begin{equation}
\frac{h}{m}=\frac{2 \pi |\delta_{sel}-\delta_{meas}|}{2 N k_B(k_1+k_2)},
\end{equation}
where $h$ is the Planck constant.
More importantly, the fine structure constant $\alpha$ can be deduced from the value of $h/m$ with the well-known relation
\begin{equation}
\alpha^2=\frac{2 R_{\infty}}{c}\frac{m}{m_e}\frac{h}{m},
\label{alpha}
\end{equation}
where $R_{\infty}$ is the Rydberg constant, $c$ is the speed of light, and ${m_e}$ is the electron mass.

The best measurement of $h/m$ was given in Ref. \cite{clade2011}, where the 5S$_{1/2}$ $|$F=2, m$_F$=0$\rangle$ and 5S$_{1/2}$ $|$F=1, m$_F$=0$\rangle$ states of $^{87}$Rb atoms were adopted as $|g\rangle$ and $|e\rangle$, respectively. The atomic beam was prepared in the $F=2$ state with an initial velocity $v_i=20$ m/s. The Bloch oscillations were generated by a Ti:sapphire laser with wavelength $\lambda_B=2\pi/k_B=532$ nm. A number of $N=500$ Bloch oscillations were applied to $^{87}$Rb atoms in each run. After a careful analysis of error budget, the value of $h/m_{Rb}$ was measured to be
\begin{equation}
\frac{h}{m_{Rb}}=4.5913592729(57)\times 10^{-9} m^2 s^{-1}.
\label{hmeas}
\end{equation}

In the next section, we will discuss how to use the above measurement to constrain parameters in previous three GUP proposals.

\section{bounds on GUP parameters}

\subsection{Bound on the KMM proposal}

Following the procedure in Ref. \cite{das2008}, we make redefinitions in the KMM proposal (\ref{kmm}),
\begin{equation}
x_i=x_{0i}, \,\,\,\,\,\,\, p_i=p_{0i}\left(1+\frac{\beta_0}{(M_P c)^2} p_0^2\right),
\label{redef}
\end{equation}
where $p_0^2=\sum_{j=1}^3 p_0^j p_{0j}$. It is easy to check that $x_{0i}$ and $p_{0j}$ now satisfy the canonical commutation relations $[x_{0i}, p_{0j}]=i \hbar \, \delta_{ij}$. Thus, it is natural to interpret $p_{0i}$ as momentum at a low-energy scale, i.e., $p_{0i}=-i\hbar d/dx_{0i}$ in position space. $p_i$ can be regarded as momentum at a high-energy scale.

Consider a quantum system with the Hamiltonian
\begin{equation}
H = \frac{p^2}{2 m} + U(\vec{x}).
\label{hamiltonian}
\end{equation}
With the redefinitions (\ref{redef}), the above Hamiltonian can be rewritten as
\begin{equation}
H = H_0 + H_1 + \mathcal{O}(\beta_0^2/(M_P c)^4),
\label{gupkin}
\end{equation}
where
\begin{equation}
H_0= \frac{p_0^2}{2 m} + U(\vec{x}), \,\,\,\, {\rm and} \,\,\,\, H_1= \frac{\beta_0}{(M_P c)^2 m} p_0^4 .
\end{equation}
Higher order terms are omitted.

Next we apply the GUP-corrected Hamiltonian [Eq. (\ref{gupkin})] to the cold $^{87}$Rb atoms. It turns out that the GUP effects modify the kinetic energy of $^{87}$Rb atoms into
\begin{equation}
E_{kin}(p_0)=\frac{p_0^2}{2 m_{Rb}}+ \frac{\beta_0}{(M_P c)^2 m_{Rb}} p_0^4.
\label{kinnew1}
\end{equation}
Repeating the previous derivation, we find a GUP-corrected relation:
\begin{eqnarray}
\nonumber |\delta_{sel}-\delta_{meas}| & &= \frac{\hbar}{m_{Rb}} 2 N k_B (k_1 + k_2) \\
\times & & \left(1+ \frac{4 m_{Rb}^2 \beta_0}{M_P^2 c^2}(v_f^2 + v_f v_i + v_i^2)\right).
\end{eqnarray}
Combining this equation with Eq. (\ref{alpha}), one can get
\begin{eqnarray}
\nonumber \frac{2 \pi |\delta_{sel}-\delta_{meas}|}{2 N k_B(k_1+k_2)}\left(1- \frac{4 m_{Rb}^2 \beta_0}{M_P^2 c^2}(v_f^2 + v_f v_i + v_i^2)\right)& &\\
=\frac{\alpha^2 c}{2 R_{\infty}}\frac{m_e}{m_u}\frac{m_u}{m_{Rb}},\,\,\,\,\,\,\,\,\,\, & &
\label{h1}
\end{eqnarray}
where $m_u$ is the atomic mass unit. Note that $R_{\infty}$, $\alpha$, $m_e/m_u$, and $m_{Rb}/m_u$ are all measured with very high accuracy.

With values in Table I, and the measurement (\ref{hmeas}), we finally find that
\begin{equation}
\beta_0 < 1.3 \times 10^{39}.
\end{equation}
This bound on $\beta_0$ is weaker than those set by high-energy physics and measurements of the Lamb shift, and better than the one from measurements of Landau levels \cite{das2008}. The reason is that $\beta_0$ is always associated with the factors, $m^2/M_P^2$ and $v^2/c^2$, in the calculation. Even if heavier atoms are chosen, the factor $m^2/M_P^2$ could not be substantially greater than $10^{-34}$. Since we are talking about cold atoms, the factor $v^2/c^2$ is at most of order $10^{-14}$. Taking these two limitations into consideration, it is hopeless for cold atom recoil experiments to give a better bound on $\beta_0$ than the one from high-energy physics. But things are different for the ADV proposal.

\begin{table}
\caption{Quantities used in our calculation}
\begin{tabular}{|l|l|c|c|}
  \hline
  Quantity & Value & Precision & Source \\ \hline
  $\alpha^{-1}$ & 137.035999173(35) & $2.5\times 10^{-10}$ & \cite{hanneke2008} \\
  $R_{\infty}$ & 10973731.568539(55) $m^{-1}$ & $5.0\times 10^{-12}$ & \cite{codata2010} \\
  $m_{Rb} $ & 86.909180535(10) $ m_u$ & $1.2\times 10^{-10}$ & \cite{mount2010} \\
 $  m_e $ & $ 5.4857990946(22)\times 10^{-4} m_u$ & $4.0\times 10^{-10}$ & \cite{codata2010} \\
  \hline
\end{tabular}
\end{table}

\subsection{Bound on the ADV proposal}

Similar to the case of the KMM proposal, make redefinitions in the ADV proposal:
\begin{equation}
x_i=x_{0i}, \,\,\,\, \, p_i=p_{0i}\left(1-\frac{\eta_0}{M_P c} p_0+\frac{2\eta_0^2}{(M_P c)^2} p_0^2\right).
\label{ADVred}
\end{equation}
Now $x_{0i}$ and $p_{0j}$ satisfy the canonical commutation relations $[x_{0i}, p_{0j}]=i \hbar \, \delta_{ij}$.
Then, taking the redefinitions (\ref{ADVred}) to the Hamiltonian Eq. (\ref{hamiltonian}), we have
\begin{equation}
H =\frac{p_0^2}{2 m}- \frac{\eta_0}{m M_P c} p_0^3+ \frac{5 \eta_0^2}{2m (M_P c)2} p_0^4 + U(\vec{x}).
\end{equation}

We the apply this GUP-corrected Hamiltonian to the cold $^{87}$Rb-atom system. The kinetic energy of $^{87}$Rb atoms is modified by the GUP effects as
\begin{equation}
E_{kin}=\frac{p_0^2}{2 m_{Rb}}- \frac{\eta_0}{m_{Rb} M_P c} p_0^3+ \mathcal{O}(\eta_0^2/(M_P c)^2),
\end{equation}
where only the linear term in $\eta_0/(M_P c)$ is kept.
By a similar derivation, one can get a GUP-modified relation
\begin{eqnarray}
\nonumber |\delta_{sel}-\delta_{meas}| & &= \frac{\hbar}{m_{Rb}} 2 N k_B (k_1 + k_2) \\
\times & & \left(1- \frac{3 m_{Rb} \eta_0}{M_P c}(v_f+ v_i)\right).
\end{eqnarray}
Using Eq. (\ref{alpha}), we find that
\begin{eqnarray}
\nonumber \frac{2 \pi |\delta_{sel}-\delta_{meas}|}{2 N k_B(k_1+k_2)}\left(1+\frac{3 m_{Rb} \eta_0}{M_P c}(v_f+ v_i)\right)& &\\
=\frac{\alpha^2 c}{2 R_{\infty}}\frac{m_e}{m_u}\frac{m_u}{m_{Rb}}.\,\,\,\,\,\,\,\,\,\, & &
\label{h2}
\end{eqnarray}

The measurement (\ref{hmeas}) and quantities in Table I determine an upper bound
\begin{equation}
\eta_0 < 2.4 \times 10^{14}.
\end{equation}
This bound on $\eta_0$ is better than those set by high-energy physics and measurements of Landau levels, and weaker than the one from measurements of the Lamb shift \cite{das2011}. Note that $\eta_0$ is always associated with the factors, $m/M_P$ and $v/c$, in the calculation. Compared to the case of the KMM proposal, effects from the ADV proposal are larger. Better bounds can be found in the future. But it is challenging to set an order-of-unity bound on $\eta_0$ with cold atom recoil experiments.

\subsection{Bound on Maggiore's proposal}

As before, we make the following redefinitions in Maggiore's proposal
\begin{equation}
x_i=x_{0i}, \,\,\,\,\,\,\, p_i=p_{0i}\sqrt{1+\frac{\gamma_0}{(M_P c)^2}(p^2+m^2 c^2)}.
\label{Maggiorered}
\end{equation}
Then, $x_{0i}$ and $p_{0j}$ satisfy the canonical commutation relations $[x_{0i}, p_{0j}]=i \hbar \, \delta_{ij}$.
If we apply the redefinitions (\ref{Maggiorered}) to the Hamiltonian Eq. (\ref{hamiltonian}), we can get
\begin{equation}
H=\frac{p_0^2}{2 m}+ \frac{\gamma_0 m}{2 M_P^2} p_0^2+ \frac{\gamma_0}{2m (M_P c)^2} p_0^4+U(\vec{x}).
\end{equation}

With this GUP-corrected Hamiltonian, the kinetic energy of $^{87}$Rb atoms is consequently modified into
\begin{equation}
E_{kin}=\frac{p_0^2}{2 m_{Rb}}+ \frac{\gamma_0 m_{Rb}}{2M_P^2} p_0^2+ \frac{\gamma_0}{2m_{Rb} (M_P c)^2} p_0^4.
\end{equation}
After a simple derivation, we can find a GUP-modified relation
\begin{eqnarray}
\nonumber & & |\delta_{sel}-\delta_{meas}| = \frac{\hbar}{m_{Rb}} 2 N k_B (k_1 + k_2) \\
& &  \,\,\,\,\,\, \times\left(1+\frac{m_{Rb}^2 \gamma_0}{M_P^2}\left(1+\frac{2(v_f^2+v_f v_i+v_i^2)}{c^2}\right)\right).
\end{eqnarray}
Together with Eq. (\ref{alpha}), we finally get
\begin{equation}
\frac{2 \pi |\delta_{sel}-\delta_{meas}|}{2 N k_B(k_1+k_2)}\left(1-\frac{m_{Rb}^2 \gamma_0}{M_P^2}\right)
=\frac{\alpha^2 c}{2 R_{\infty}}\frac{m_e}{m_u}\frac{m_u}{m_{Rb}},
\label{h3}
\end{equation}
where terms involved in $v_i$ and $v_f$ have been omitted.

Thus, the measurement (\ref{hmeas}), together with quantities in Table I, sets an upper bound
\begin{equation}
\gamma_0 < 1.1\times 10^{26}.
\end{equation}
It is obvious that this bound on $\gamma_0$ is much better than the one set by high-energy physics, which is very different from the case for the KMM proposal. The reason is that $\gamma_0$ is only associated with the factor $m^2/M_P^2$ in the calculation. Without a further suppression due to the factor $v^2/c^2$, the bound on $\gamma_0$ is almost 13 orders better than the one on $\beta_0$. This feature makes Maggiore's proposal easier to test than the KMM proposal.

\section{Conclusion}

We have discussed three proposals for the GUP, and investigated how to use results from cold atom recoil experiments to constrain parameters in these proposals. The main reason to take the $^{87}$Rb atom recoil experiment is that all the physical quantities involved can be measured with very high precision. Compared to bounds set from high-energy physics and other experiments, our bound on the KMM proposal is worse, which is mainly due to the suppression by the factor $v^2/c^2$. In other words, cold atom recoil experiments are not suitable for studying effects of the KMM proposal. Things are totally different for the ADV proposal and Maggiore's proposal. We get the second best bound on the ADV proposal, and the best bound on Maggiore's proposal. The high preciseness of the $^{87}$Rb atom recoil experiment plays the crucial role in giving these impressive bounds. However, these bounds are still many orders higher than theoretical predictions. Thus, to test various theories of quantum gravity, we have to resort to other possible atomic physics experiments.

\begin{center}
\large{{\bf Acknowledgements}}
\end{center}

This work was supported by the National Key Research Program of China under Grant No. 2016YFA0302002, the National Science Foundation of China under Grants No. 11227803 and No. 91536221, and by funds from the Chinese Academy of Sciences.

\end{document}